\documentclass[fleqn,10pt]{wlscirep}
\usepackage[utf8]{inputenc}
\usepackage[T1]{fontenc}
\usepackage{textalpha}
\title{DEEP\textsuperscript{2}: \underline{Deep} Learning Powered \underline{De}-scattering with \underline{E}xcitation \underline{P}atterning}

\author[1,2]{Navodini Wijethilake}
\author[1,2]{Mithunjha Anandakumar}
\author[3,4]{Cheng Zheng}

\author[3,4,6]{Peter T. C. So}
\author[4,5,7]{Murat Yildirim}
\author[1,*]{Dushan N. Wadduwage}
\affil[1]{Center for Advanced Imaging, Faculty of Arts and Sciences,  Harvard University, Cambridge, USA}
\affil[2]{Department of Electronic and Telecommunication Engineering, University of Moratuwa, Sri Lanka}
\affil[3]{Department of Mechanical Engineering, Massachusetts Institute of Technology, 77 Massachusetts Ave., Cambridge, MA 02139, USA.}
\affil[4]{Laser Biomedical Research Center, Massachusetts Institute of Technology, 77 Massachusetts Ave., Cambridge, MA 02139, USA.}
\affil[5]{Picower Institute for Learning and Memory, Massachusetts Institute of Technology, 77 Massachusetts Ave., Cambridge, MA 02139, USA.}
\affil[6]{Department of Biological Engineering, Massachusetts Institute of Technology, 77 Massachusetts Ave., Cambridge, MA 02139, USA.}
\affil[7]{Department of Neuroscience, Cleveland Clinic Lerner Research Institute, Cleveland, OH 44195, USA.}

\affil[*]{\href{mailto:wadduwage@fas.harvard.edu}{wadduwage@fas.harvard.edu}}

\begin{abstract}
\noindent Limited throughput is a key challenge in in-vivo deep-tissue imaging using nonlinear optical microscopy. Point scanning multiphoton microscopy, the current gold standard, is slow especially compared to the wide-field imaging modalities used for optically cleared or thin specimens. We recently introduced “De-scattering with Excitation Patterning or DEEP”, as a widefield alternative to point-scanning geometries. Using patterned multiphoton excitation, DEEP encodes spatial information inside tissue before scattering. However, to de-scatter at typical depths, hundreds of such patterned excitations are needed. In this work, we present DEEP$^2$, a deep learning based model, that can de-scatter images from just tens of patterned excitations instead of hundreds. Consequently, we improve DEEP’s throughput by almost an order of magnitude. We demonstrate our method in multiple numerical and physical experiments including in-vivo cortical vasculature imaging up to four scattering lengths deep, in alive mice.
\end{abstract}
\begin{document}

\flushbottom
\maketitle
%
%
\thispagestyle{empty}


\section*{Introduction}
\noindent Imaging biological structures such as neurons or blood vessels deep inside scattering tissue is an important yet challenging microscopy problem. For such in-vivo deep tissue experiments, today, the gold standard is the point-scanning two- (or three-) photon microscope (PSTPM) \cite{rocheleau2003two,yildirim2019functional,yildirim2022label}. Especially, PSTPM occupies a unique position in neuroscience, due to its high spatial resolution, low phototoxicity, and ability to penetrate through deep tissue. In operation, PSTPM focuses femtosecond laser light at long wavelengths through scattering tissue to excite fluorescent molecules. Due to long wavelengths, the excitation light sees little scattering. The microscope then collects emission fluorescence through the same scattering tissue onto a detector. This emission light however is at a much shorter wavelength, and hence encounters significant scattering on its way to the detector. For this reason, multiple points cannot be excited and resolved at the same time. So imaging is performed sequentially one point at a time. Thus, PSTPM is inherently a slow imaging technique. When speed is needed, either resolution or the imaging field of view should be compromised \cite{benninger2013two}.\\

\noindent Despite the limitations due to emission light scattering, wide-field two- and three-photon microscopes (WFTPM) have been proposed. For instance, temporal focusing microscopy (TFM) focuses amplified femtosecond laser pulses temporally, for depth selective WFTPM. TFM provides excellent optical sectioning and also penetrates well through scattering tissue due to long multi-photon excitation wavelengths~\cite{vaziri2010ultrafast, rowlands2017wide}. Yet, the emission light scatters. Therefore, in wide-field deep-tissue imaging, some photons are mapped onto incorrect detector pixels, degrading the SNR and spatial resolution. Few groups, including us, recently proposed to overcome this limitation by combining TFM with structured light illumination. Since excitation light penetrates through the scattering medium, structured illuminations can modulate the imaging field of view before scattering. Thus, the information about the structured illuminations can be used to de-scatter TFM images despite degraded detection. Using this principle, Escobet et al. \cite{escobet2018wide} proposed “TempoRAl Focusing microscopy with single-pIXel detection (TRAFIX)”. They used coded excitations and single-pixel detection. We proposed ‘De-scattering by Excitation Patterning (DEEP)’ \cite{wadduwage2019scattering}. Instead of single-pixel detection, we used wide-field detection. Single-pixel detection completely relies on coded illuminations for de-scattering. Wide-field detection, however, retains some spatial information and only partially depends on coded illumination. Thus, while TRAFIX needs tens of thousands of illumination patterns to reconstruct a typical WFTPM image, DEEP requires only hundreds of illumination patterns. Despite such speed improvement, DEEP is still orders of magnitude slower  compared to other depth-resolved wide-field imaging modalities used for optically cleared or thin specimens. For instance, modalities such as single plane illuminated light-sheet microscopy (SPIM) and structured-illumination microscopy (SIM) are either single shot or few-shot acquisition. Therefore, in this work, we attempt to further increase the throughput of DEEP by utilizing prior image information within a deep-learning-based computational imaging framework.\\

\noindent From a computational imaging point of view, the imaging process that describes the translation of the ideal image $(x)$ to the observed image $(y)$ is called the forward model $(f)$. During the forward imaging process, the observed image $(y)$ is degraded by image noise, low-pass filtering, pixel-value quantization, sub-sampling, and in our case scattering \cite{belthangady2019applications}. An inverse model should be constructed to map the observed $(y)$ to the ideal expected image $(x)$. In our original work~\cite{wadduwage2019scattering}, we used an analytical inverse model that utilized wavelet sparsity priors. However, recently, for such inverse problems, deep learning has proven impressively capable. \cite{belthangady2019applications,weigert2018content,jin2017deep}. For example, deep learning has shown outstanding success in image classification \cite{krizhevsky2012imagenet,ziletti2018insightful}, biomedical imaging \cite{wei2019dominant,liu2014early,wang2016accelerating}, segmentation \cite{girshick2014rich}, prediction \cite{nielsen2018deep}, denoising \cite{eraslan2019single} and other linear and non-linear inverse problems \cite{jin2017deep,eulenberg2017reconstructing,wei2019physics,zhu2018image}. In our previous work we demonstrated that deep learning can reconstruct fine biological structures such as dendritic spines from scattered TFM images \cite{wei20193d}. Similarly, in this work we present a deep learning-based inverse reconstruction method for DEEP-TFM.\\

\noindent Our proposed method is termed DEEP$^2$. DEEP$^2$ consists of a learning-powered inverse model that can reconstruct a de-scattered image from only 32-DEEP measurements (instead of 256 in our original work). The model architecture is inspired by the UNet but modified to include a self-attention mechanism on the expanding path and a terminating reconstruction block. We trained DEEP$^2$ using simulated data sets. To simulate training data, we first carefully modeled the forward imaging process of the DEEP microscope. As a part of the forward model, we also modeled the optical behavior of scattering tissue using Monte Carlo simulations. We then used the forward model to simulate DEEP images from PSTPM-like image stacks (simulated or experimentally acquired). The DEEP$^2$, trained on the simulated data was finally tested on the experimentally acquired DEEP images of beads and mouse cortical vasculature. Our results suggest that the proposed method can reconstruct deep tissue images up to four scattering lengths deep in mouse cortical vasculature, using only tens of patterned excitations instead of hundreds.

\section*{Results and Discussion}

\begin{figure}[!ht]
\centering
\includegraphics[width=0.6\linewidth]{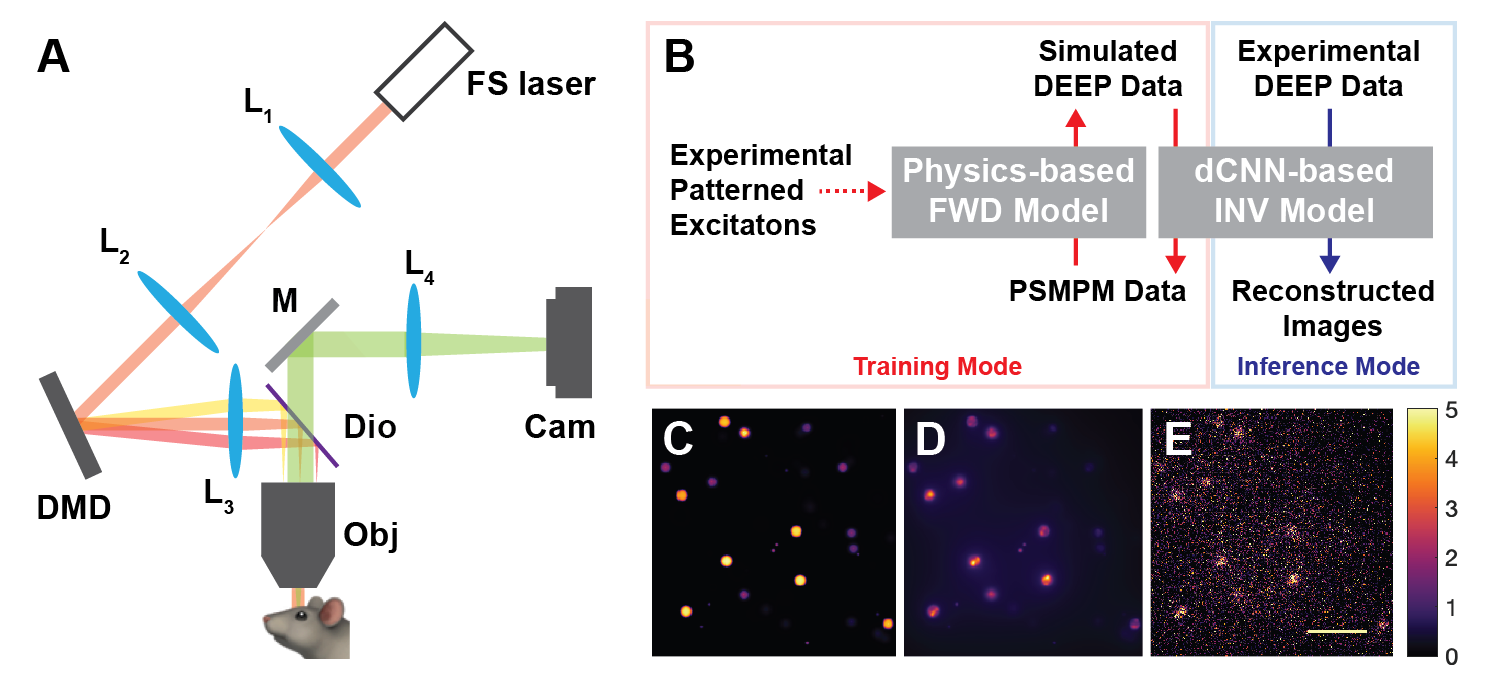}
\caption{The proposed deep learning based DEEP imaging method. \textbf{(A)} Optical schematic of the DEEP-TFM microscope: FS laser - Amplified femtosecond laser; L1 \& L2 - Optical relay; DMD - Digital micro-mirror device; L3 - Excitation tube lens; Dio - Dichroic mirror; Obj - Objective lens; M - Mirror; L4 - Emission tube lens; Cam - Camera Detector. \textbf{(B)} Schematic of the proposed physics based forward model and the deep learning based 'DEEP$^2$' inverse model. \textbf{(C)} A representative simulated object. \textbf{(D)} Image of the object in 'C' at the final image plane of the microscope. \textbf{(E)} The image in 'D' after detected using the EMCCD camera. Note the low photon count and resulting noise.}
\label{fig:main_method}
\end{figure}

\subsection*{Deep-learning Powered DEEP Microscopy}
In this work, we propose a deep-learning-based inverse model to reconstruct images from DEEP (De-scattering with Excitation Patterning) measurements. Fig.\ref{fig:main_method}A shows the optical schematic of the DEEP-TFM microscope. Amplified femtosecond laser pulses are temporally focused for depth-resolved multi-photon excitation. In the excitation path, a digital micro-mirror device (DMD) is placed at a conjugate image plane before the excitation tube lens. The DMD mirrors project a binary patterned excitation onto the sample at the focal plane. The emission photons are collected by the objective lens and imaged onto a camera detector. During the experiment, multiple patterns are used to generate multiple encoded image measurements. Our goal is to learn an inverse model to reconstruct de-scattered images using the knowledge about the excitation patterns and measured images (see Fig.\ref{fig:main_method}B). To train such a model, we first generated simulated training data using a physics-based forward image model of the DEEP-TFM microscope and the scattering process. We discuss the forward model, the EMCCD detector model, and the scattering model in detail in the methods section. Excitation patterns are parameters of the said forward model (see Fig.\ref{fig:main_method}B). We first measured these patterns in a calibration experiment. Next, PSTPM-like ground-truth images were input to the forward model to simulate DEEP images (see Fig.\ref{fig:main_method}C-E). This allowed us to generate paired DEEP and PSTPM data to train the inverse model. The inverse model was based on the seminal UNet architecture. We further modified the UNet by adding an attention mechanism to the expanding path. Details about the inverse model are presented in the method section. 

To test our method, we experimented on three datasets: (1) a mixture of fluorescent beads, (2) mouse pyramidal neurons with the dendritic arbor, and (3) mouse cortical vasculature. We synthetically generated PSTPM-like beads images for training (see methods). Mouse pyramidal neurons and cortical vasculature were imaged from multiple animals using PSTPM to generate the training datasets. We then performed physical DEEP experiments on a mixture of fluorescent beads and cortical vasculature of anesthetized mice (see methods) and used our inverse model (trained on simulated data) to successfully de-scatter experimental DEEP measurements.

In the next two sections, we present numerical reconstruction results on unseen test data sets for each case; and de-scattering results for experimental DEEP measurements.

\begin{figure}[!h]
\centering
\includegraphics[width=0.8\linewidth]{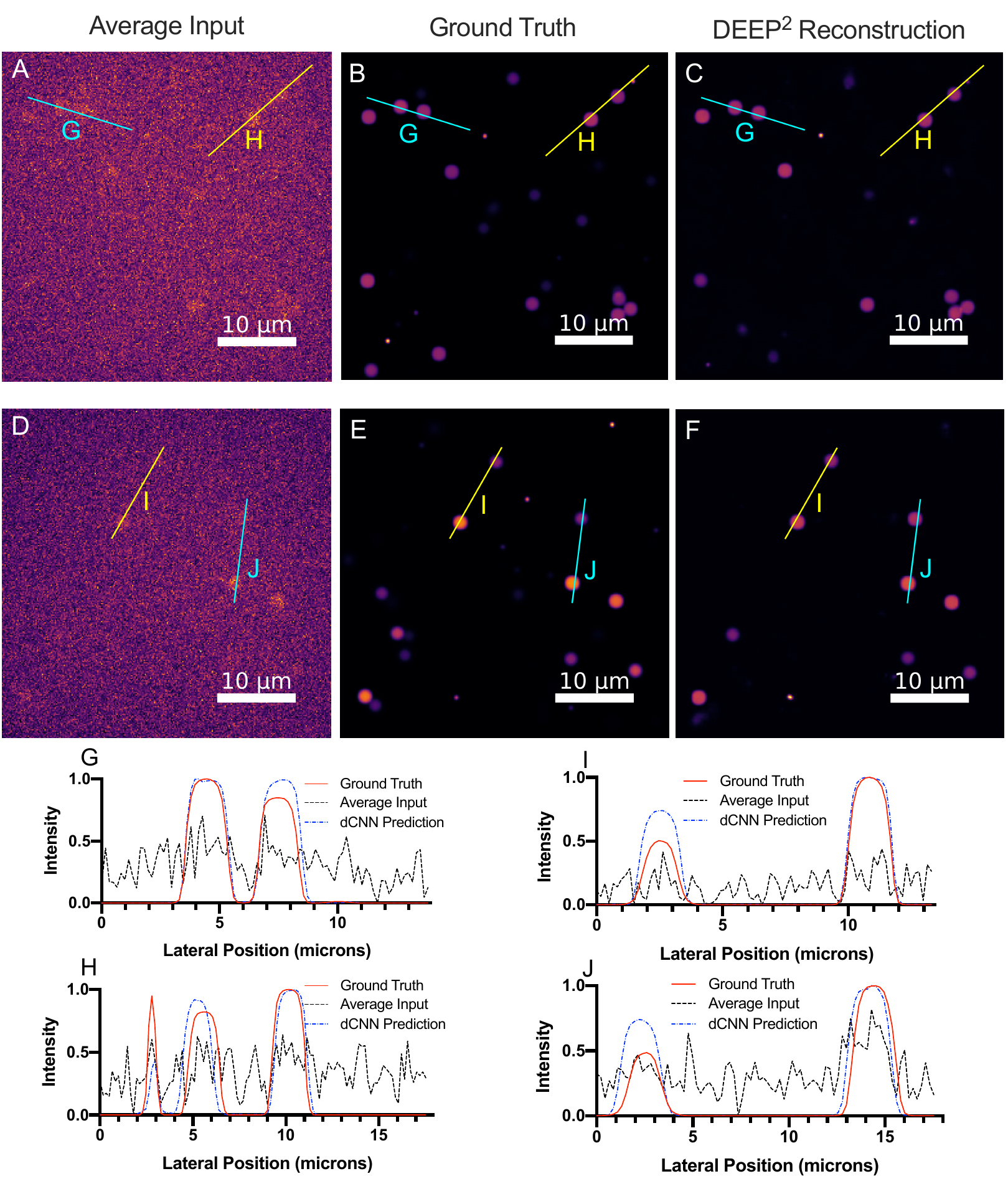}
\caption{DEEP$^2$ Validation results on the Synthetic Fluorescent Beads. The beads are in the size of 1.5 um and 6 um, with the intensity of the 1.5 um beads 5X higher than the 6 um beads.  \textbf{(A)} \& \textbf{(D)} Two synthetic DEEP-TFM instances generated using the forward model. Only one patterned image, out of 32 patterned images used as the input for the DEEP$^2$, is visualized. \textbf{(B)} \& \textbf{(E)} Corresponding synthetic ground truths for the (A) \& (B) instances. \textbf{(C)} \& \textbf{(F)} DEEP$^2$ reconstruction using the 32 patterned DEEP-TFM synthetic images corresponding to (A) \& (D) instances. The intensity along the lines G and H, shown in blue and yellow on (A) is visualized in \textbf{(G)} \& \textbf{(H)}. Similarly, the intensity along the lines I and J are shown in in \textbf{(I)} \& \textbf{(J)} plots. \textbf{(K)} shows the variance of the 32 patterns. }
\label{fig:beads_test}
\end{figure}

\subsection*{Numerical Results on Synthetic Test Data}
In this section we present numerical results on synthetic test datasets to validate our trained inverse model. All experiments used only 32 patterned illuminations which is nearly an order of magnitude less than the current state of the art \cite{wadduwage2019scattering}.


\paragraph{Reconstruction Results on Synthetic Beads Data.} Simulated 3D bead stacks were used as the input to the forward model to obtain the synthetic DEEP-TFM like instances at 5 scattering lengths. Then, the DEEP$^2$ based inverse model was trained on the synthetic instances and was validated on similarly generated unseen synthetic beads instances. Fig.\ref{fig:beads_test} presents two validated instances, where the average input is shown in the first column, the ground truth in the second column, and the reconstruction of the DEEP$^2$ in the third column. In addition, the intensities along the lateral directions (G, H, I J in Fig.\ref{fig:beads_test}), were plotted for further clarification. The DEEP$^2$ successfully reconstructed the small beads with a radius of 1.5 um. 

To obtain the best performing DEEP$^2$, we employed several approaches.
First, we compared the vanilla UNet and the UNet with the scSE module, along with the KL divergence loss, to obtain the architecture-wise best inverse model. We could clearly observe that, in the validation, SSIM value was $> 0.9$ for the modified UNet and $> 0.8$ for the vanilla UNet.  Further, we exploited 4 different loss functions on the modified UNet. We clearly demonstrated that the SSIM values for the MSE, RMSLE, and Smooth L1 losses were numerically low compared to the KL Divergence loss. Additionally, for the validation instances, the SSIM, MSE, and PSNR metric distributions were statistically assessed. For this, Kruskal Wallis nonparametric test was used to examine whether each metric is similarly distributed across each loss function. Thus, the KL Divergence loss was significantly different ($p < 0.05$) from the other 3 loss functions for all three metrics.

\begin{figure}[!h]
\centering
\includegraphics[width=0.9\linewidth]{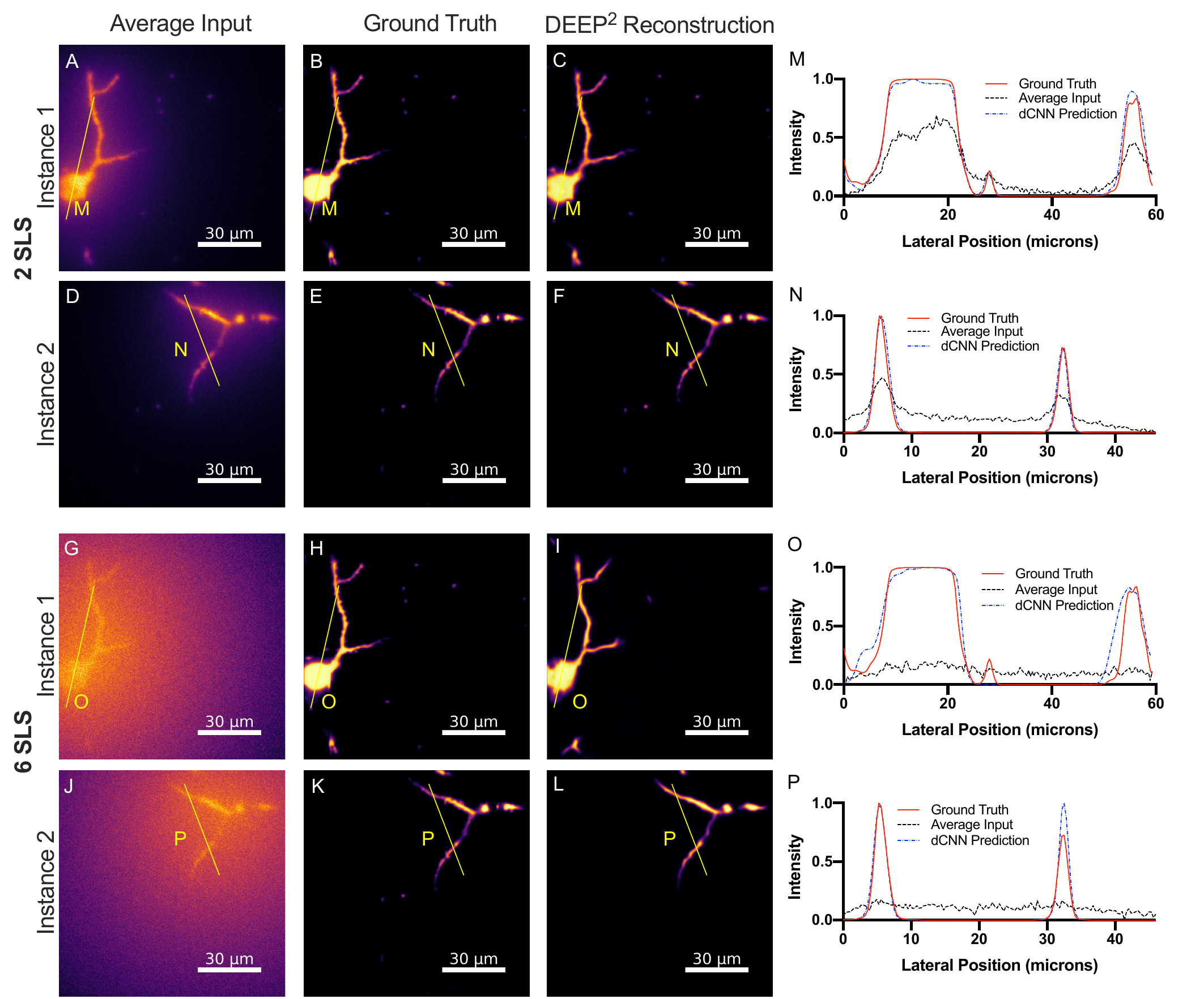}
\caption{DEEP$^2$ validation results on the mouse pyramidal neurons with the dendritic arbor for 2 and 6 scattering length depths. \textbf{(A)} Average of the 32 patterned DEEP-TFM images, generated using the forward model for a single instance at 2 SLS. \textbf{(B)} The ground truth for that instance, a PSTPM image. \textbf{(C)} The descattered instance using the DEEP$^2$ with the 32 patterned DEEP-TFM images at 2 SLS. \textbf{(D)-(F)} Another instance at 2 SLS. (E) is a PSTPM image used as the ground truth, (F) is the DEEP$^2$ reconstruction.}
\label{fig:neuron_main}
\end{figure}

\paragraph{Mouse Pyramidal Neuron Reconstructions.} Figure. \ref{fig:neuron_main} shows the validation cohort results on the mouse pyramidal neurons with a dendritic arbor. We have presented two instances corresponding to 2 and 6 scattering length depths and the respective intensities along the lateral directions on each instance are presented in the last column. Figure. \ref{fig:neuron_main} (A) and (D) are two separate instances, each representing average of 32 patterned DEEP-TFM images generated using the forward model with the PSTPM mouse pyramidal neuron images at 2 scattering lengths. Figure. \ref{fig:neuron_main} (B) and (E), represent the PSTPM images, which were used as the ground truth for the corresponding (A) and (D) instances. Figure. \ref{fig:neuron_main} (C) and (F) are the reconstructions of DEEP$^2$, that de-scattered the patterned 32 DEEP-TFM images at 2 scattering lengths. The intensity along the yellow line on (A)-(C) and (D)-(F) are visualized in (M) and (N) graphs respectively.  Similarly, Figure. \ref{fig:neuron_main} (G) and (J) show an average of 32 patterned DEEP-TFM images, synthesized using the forward model, for the same structures shown in (A) and (D) at 6 scattering lengths. Figure. \ref{fig:neuron_main} (H) and (K) show corresponding ground truth and (I) and (L) show the reconstruction of DEEP$^2$, at 6 scattering lengths. The intensity along the yellow line on (G)-(I) and (J)-(L) are represented on the (O) and (P) graphs respectively. Thus, we could observe at 2 SLS the DEEP$^2$ was able to reconstruct spines, fine structures along the dendrite. 

\begin{figure}[!ht] 
\centering
\includegraphics[width=1.0\linewidth]{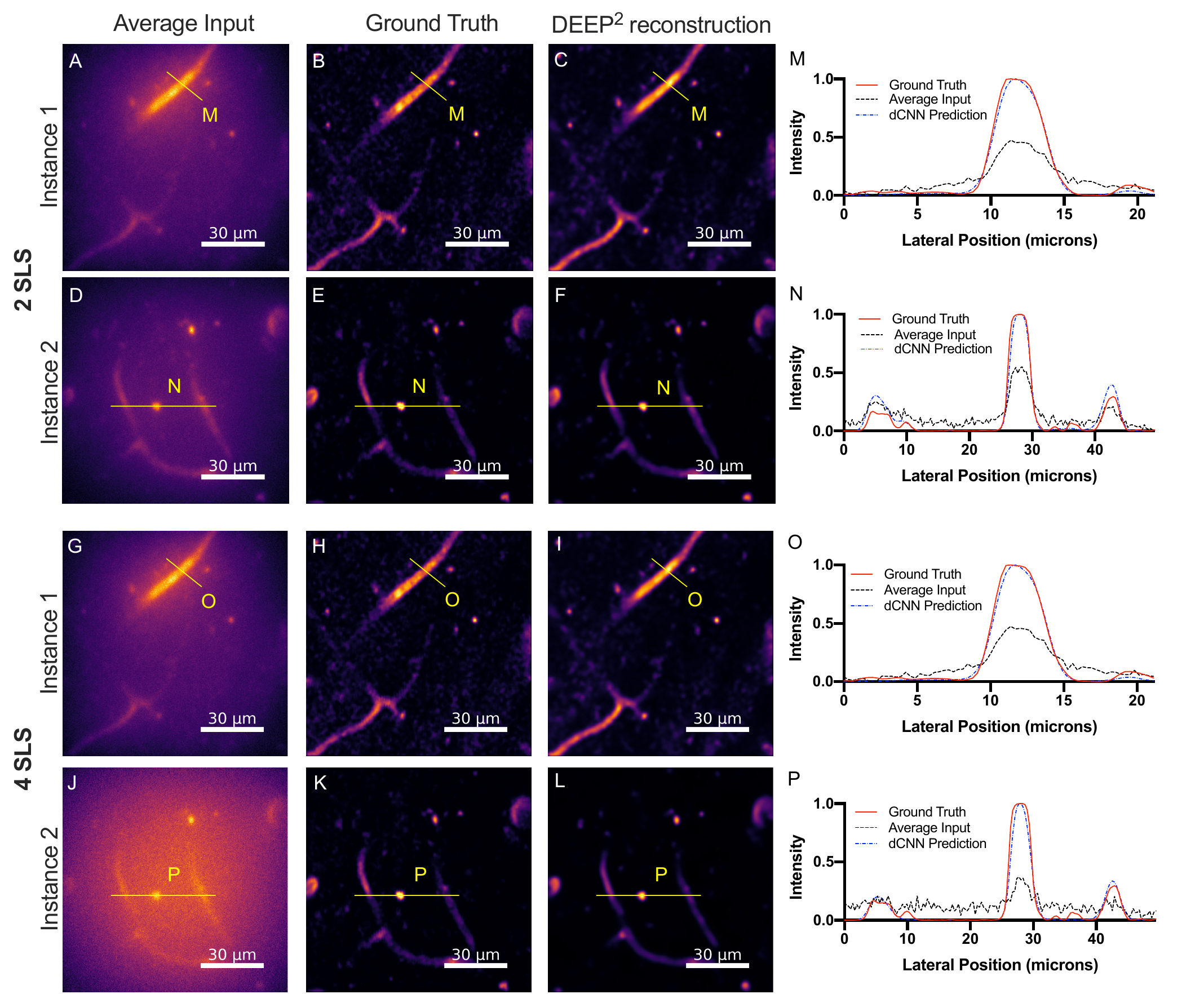}
\caption{DEEP$^2$ validation results on the mouse cortical vasculature structures at 2 and 6 scattering length depths. \textbf{(A)} Average of the 32 patterned DEEP-TFM images, generated using the forward model for a single instance at 2 SLS. \textbf{(B)} The ground truth for that instance, a three-photon Fluorescent microscope image. \textbf{(C)} The descattered instance using the DEEP$^2$ with the 32 patterned DEEP-TFM images at 2 SLS. \textbf{(D)-(F)} Another instance at 2 SLS. (E) is a three-photon Fluorescent microscope image used as the ground truth, (F) is the DEEP$^2$ reconstruction.}
\label{fig:BV_main}
\end{figure}

\paragraph{Mouse Cortical Vasculature Reconstructions.} 
The best performing model, i.e. the KL divergence loss with the modified UNet, as the DEEP$^2$, was used to reconstruct mouse cortical vasculature images from numerical DEEP data. 
Figure. \ref{fig:BV_main} shows two validation instances at 2 and 4 scattering lengths and intensity profiles along the highlighted lateral directions are presented in the last column. Figure. \ref{fig:BV_main} (A) and (D) are the average of 32 patterned DEEP-TFM images, synthesized using the forward model, for two separate mouse cortical vasculature instances. The corresponding ground truths, i.e. the three-photon Fluorescent microscope images, are shown in Figure. \ref{fig:BV_main} (B) and (E). The DEEP$^2$ reconstruction for the same two instances at 2 scattering lengths are presented in Figure. \ref{fig:BV_main} (C) and (D). The intensity profiles along the marked yellow lines on (A)-(C) and (D)-(F) are indicated in (M) and (N) graphs respectively. Similarly, for the same vasculature instances, the average over the 32 patterned DEEP-TFM instances synthesized using the forward model at the 6 scattering lengths are shown in Figure. \ref{fig:BV_main} (G) and (J). The corresponding ground truths are shown in Figure. \ref{fig:BV_main} (H) and (K) and the DEEP$^2$ reconstructions at 6 scattering lengths are shown in Figure. \ref{fig:BV_main} (I) and (L). The intensity along the lateral directions marked in yellow on (G)-(I) and (J)-(L) are represented in Figure. \ref{fig:BV_main} (O) and (P).  


\begin{figure}[h!]
\centering
\includegraphics[width=0.8\linewidth]{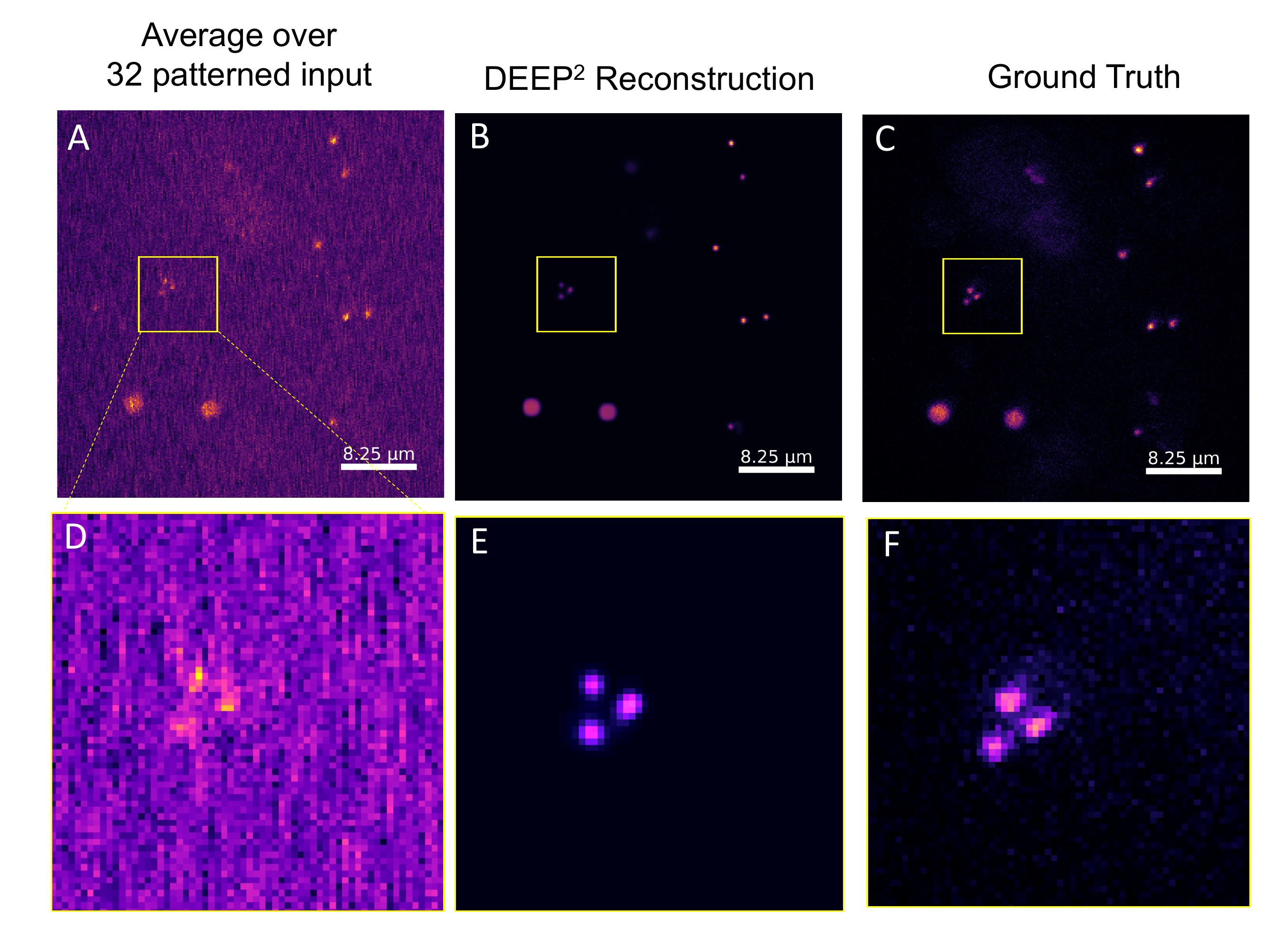}
\caption{DEEP$^2$ testing phase: qualitative evaluation on the fluorescent beads at 5 scattering lengths. \textbf{(A)} Average of the experimentally acquired 32 patterned DEEP-TFM images. \textbf{(B)} DEEP$^2$ reconstruction. \textbf{(C)} Ground truth. \textbf{(B)-(F)} The yellow colored box on (A)-(C) images are enlarged for close visualization. }
\label{fig:Exp_beads_main}
\end{figure}

\begin{figure}[h!]
\centering
\includegraphics[width=\linewidth]{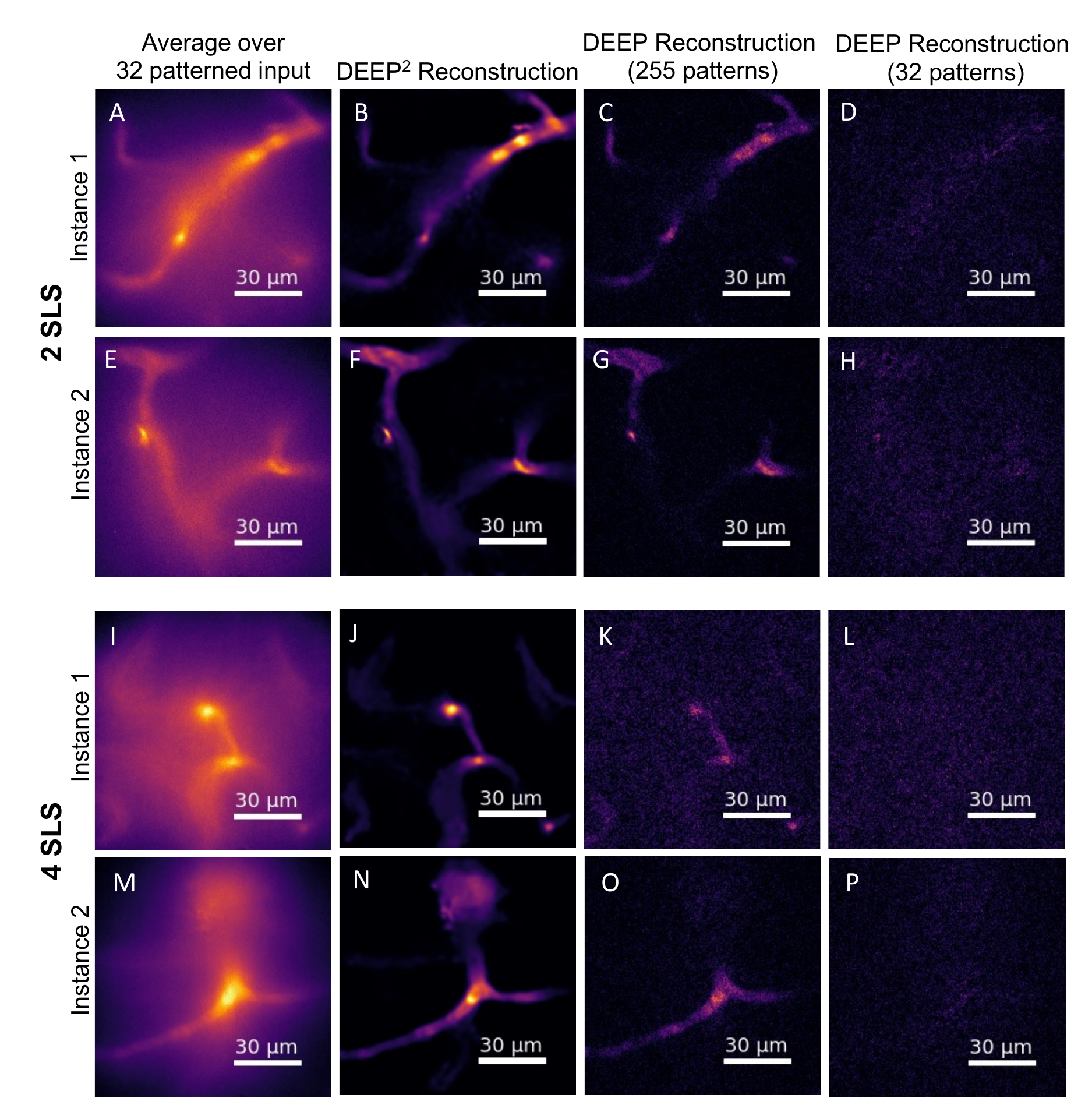}
\caption{DEEP$^2$ testing phase: qualitative evaluation on the mouse vasculature structures at 2 and 4 scattering lengths. \textbf{(A) \& (E)} Average of 32 DEEP-TFM images acquired at 2 scattering lengths. \textbf{(B) \& (F)} DEEP$^2$ reconstruction at 2 scattering lengths. \textbf{(C) \& (G)} DEEP-TFM mathematical inverse algorithm reconstruction with 255 DEEP-TFM images at 2 scattering lengths. \textbf{(D) \& (H)} DEEP-TFM mathematical inverse algorithm reconstruction with 32 DEEP-TFM images at 2 scattering lengths. \textbf{(I) \& (M)} Average of 32 DEEP-TFM images acquired at 4 scattering lengths. \textbf{(J) \& (N)} DEEP$^2$ reconstruction at 2 scattering lengths. DEEP-TFM mathematical inverse algorithm reconstruction \textbf{(K) \& (O)} with 255 DEEP-TFM images. \textbf{(L) \& (P)} with 32 DEEP-TFM images.}
\label{fig:Exp_BV_main}
\end{figure}

\subsection*{Results on Experimental Data}

In this section, we present experimental validation of DEEP$^2$ by successfully de-scattering experimentally acquired DEEP measurements using the inverse model trained on synthetic data. Experiments used only 32 patterned illuminations and thus are nearly an order of magnitude faster than the current state of the art \cite{wadduwage2019scattering}.


\paragraph{DEEP$^2$ Imaging and Reconstruction of a Mixture of Fluorescent Beads.} The DEEP$^2$ inverse model trained on the artificial beads data at 5 scattering lengths, was tested on experimentally acquired DEEP-TFM images. In accordance with the training data, similar sized beads at a similar composition were imaged through a scattering intralipid layer (resulting 5-6 scattering lengths) using the DEEP-TFM microscope with 32 patterned excitations. The images were then input to the trained-inverse model to output the reconstructions. The results are shown in Figure \ref{fig:Exp_beads_main}. Figure \ref{fig:Exp_beads_main} (A) shows the average of the 32 patterned DEEP-TFM images and (B) shows the DEEP$^2$ reconstruction. Figure \ref{fig:Exp_beads_main} (C) shows the ground truth acquired without the scattering intralipid layer. A region that consists of beads with radius of ~1 um, marked as a yellow box on Figure \ref{fig:Exp_beads_main} (A), was closely observed to evaluate the performance. As shown in the Figure \ref{fig:Exp_beads_main} (D), the average of 32 patterned DEEP-TFM images contained faint signs of those structures. But the proposed DEEP$^2$could reconstruct those fine beads, eliminating the background noise as shown in Figure \ref{fig:Exp_beads_main} (E). A quantitative evaluation was not performed as the DEEP-TFM field of view and the ground truth FOV weren’t pixel matched due to practical experimental reasons.


\paragraph{DEEP$^2$ Imaging and Reconstruction of in-vivo Mouse Cortical Vasculature.} Next, the DEEP$^2$ model trained on synthetic cortical vasculature data, was tested on experimentally acquired DEEP-TFM images of mouse cortical vasculature (see methods section for DEEP-TFM experimental details). Figure. \ref{fig:Exp_BV_main} shows several mouse cortical vascular instances at 2 and 4 scattering-length depths. Figure. \ref{fig:Exp_BV_main} (A) and (E) show the average of the 32 patterned DEEP-TFM images for two separate FOVs at 2 scattering lengths. Figure. \ref{fig:Exp_BV_main} (B) and (F) are the corresponding DEEP$^2$ reconstruction with 32 DEEP-TFM images. Figure. \ref{fig:Exp_BV_main} (C) and (G) are the DEEP-TFM mathematical inverse algorithm reconstruction with 255 DEEP-TFM images\cite{wadduwage2019scattering}. Figure. \ref{fig:Exp_BV_main} (D) and (H) are the DEEP-TFM mathematical inverse algorithm reconstruction with 32 DEEP-TFM images. Similarly, Figure. \ref{fig:Exp_BV_main} (I) and (M) show the average of the 32 patterned DEEP-TFM images for two separate FOVs at 4 scattering lengths. Figure. \ref{fig:Exp_BV_main} (J) and (N) are the DEEP$^2$ reconstructions with 32 DEEP-TFM images corresponding to (I) and (M) respectively. Figure. \ref{fig:Exp_BV_main} (K) and (O) are the DEEP-TFM mathematical inverse algorithm reconstruction with 255 DEEP-TFM images at 4 scattering lengths. Figure. \ref{fig:Exp_BV_main} (L) and (P) are the DEEP-TFM mathematical inverse algorithm reconstruction with 32 DEEP-TFM images at 4 scattering lengths. The DEEP$^2$ reconstruction (with 32 DEEP-TFM images) showed additional details that were missing in the inverse algorithm reconstructions with 255 DEEP-TFM images (compare Figure \ref{fig:Exp_beads_main} B, F, J, N to Figure \ref{fig:Exp_beads_main} C, G, K, O). However, the average of raw data (see Figure \ref{fig:Exp_beads_main} A, E, I, M) suggests that these structures are real and not false artifacts generated by the network. The inverse Algorithm failed to reconstruct with 32 DEEP-TFM images (see Figure \ref{fig:Exp_beads_main} D, H, L, P).

\section*{Conclusion}

In conclusion, we demonstrate the DEEP$^2$ for in-vivo deep tissue imaging with fine reconstruction that requires a less number of patterned instances, compared to our original work~\cite{wadduwage2019scattering}. Evaluation on the experimental instances of in-vivo cortical vasculature imaging up to four scattering lengths strengthens the applicability of DEEP$^2$ for complex biological structures. One of the limitations of this work can be identified as the high time consumption for data generation using the forward model with a comprehensive hyperparameter tuning. Thus, In future work we will focus on implementing a learning based forward model that can be integrated with the DEEP$^2$ for optimized reconstruction.

\section*{Methods}

\subsection*{Forward Model}


\begin{figure}[!ht]
\centering
\includegraphics[width=12cm]{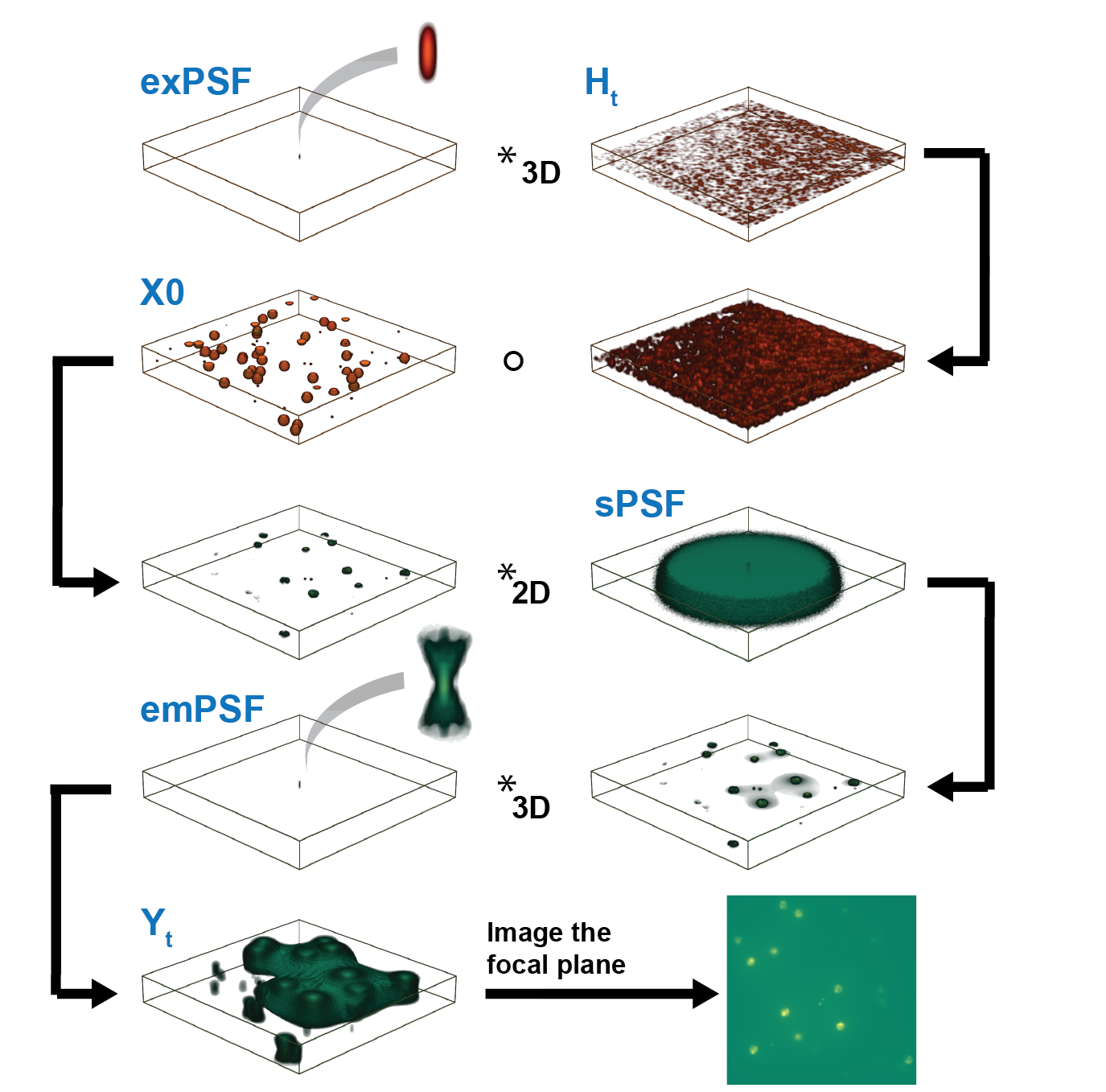}
\caption{Forward model}
\label{fig:fwd_model1}
\end{figure}

\begin{figure}[!ht]
\centering
\includegraphics[width=12cm]{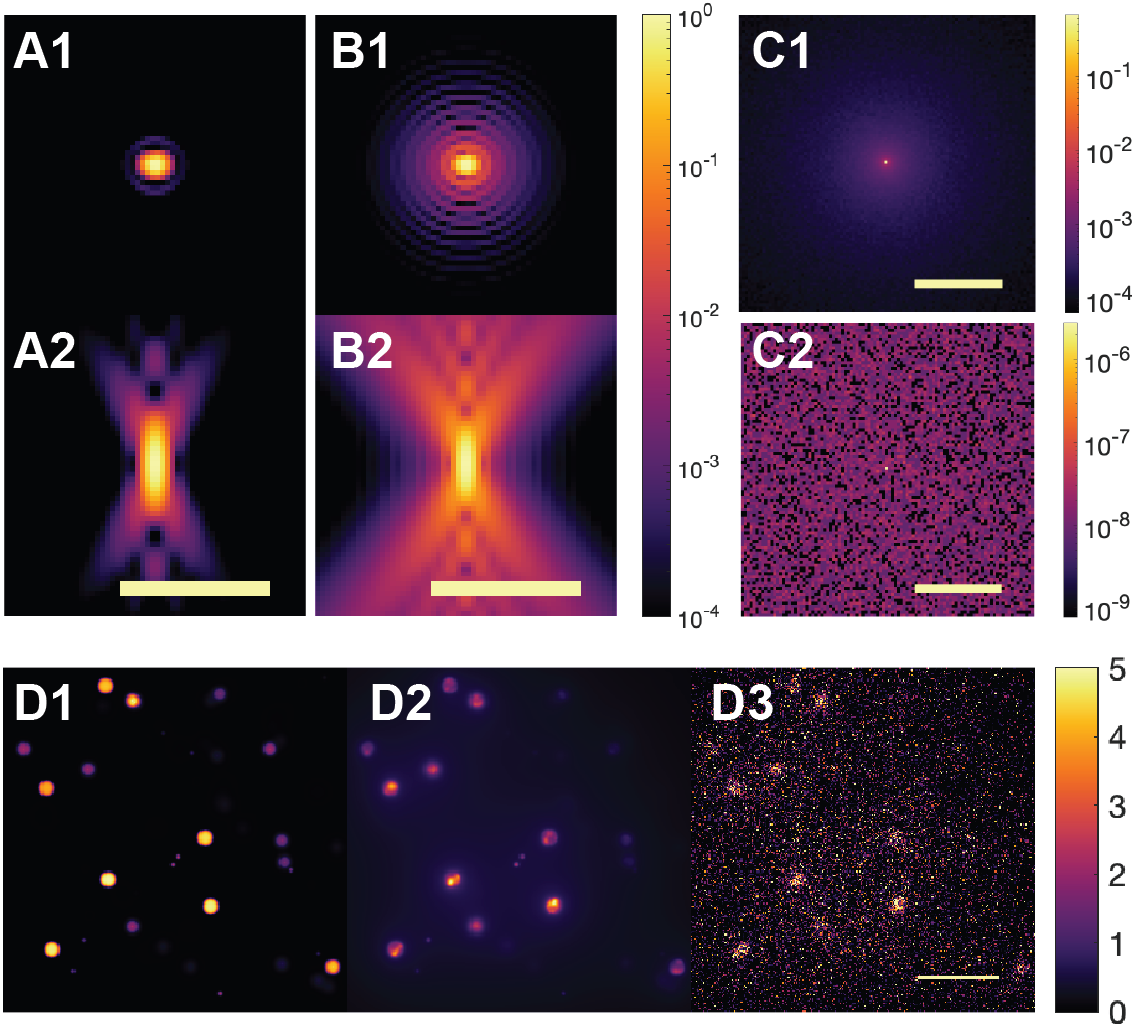}
\caption{Forward model.(A1-2) The xy and xz views of the excitation PSF. (B1-2) The xy and xz views of the emission PSF. (C1) The scattering point spread function at a two-scattering-length depth. (C2) The scattering point spread function at a seven-scattering-length depth. (D1) A simulated beads object. (D2) The simulated DEEP-TFM image of the object in 'D1' at a seven-scattering-length depth before detection. Note that the maximum photon count in the image is close to 5 photons. (D3) The simulated DEEP-TFM image in 'D2' detected on the simulated EMCCD camera. The scale bars in 'A2' and 'B2' are $5\mu m$. The scale bars in 'A2', 'B2', 'C1', and 'C2' are $5\mu m$. The scale bar in 'D3'is $20\mu m$. }
\label{fig:fwd_model2}
\end{figure}

\noindent \paragraph{Model Architecture.} To be used as inputs to our deep learning model for training, we generate DEEP-TFM image stacks synthetically from the PSTPM image stacks using a forward imaging model (see Figure \ref{fig:fwd_model1}). Let $exPSF(x,y,z)$ be the 3D excitation point spread function(PSF) and $H_t(x,y)$ be the pattern on the DMD placed at an image plane at the excitation side. Now, $exPSF(x,y)*_{3D} H_t$ is the 3D patterned excitation on the sample. Here $*_{3D}$ is the 3D convolution operation. The result is multiplied element-wise with the object, $X0(x,y,z)$ to get the excited object cube. Then the resulting 3D excited object is convolved plane-by-plane with the stack of 2D scattering point spread functions at each depth ($sPSF(x,y,z)$). This operation is represented by $*_{2D}$. Each $sPSF$ at a certain depth, represents the effective scattered light distribution from a point source at that depth. We will introduce and discuss the sPSFs model in the next section in detail. The resulting 3D cube is then convolved in 3D with the emission PSF ($emPSF(x,y,z)$). These operations constitute the following equation.
\[ Y_t(x,y,z) = \; \{ \{(exPSF(x,y,z)\;*_{3D}\; H_t(x,y))\;\circ\;X0(x,y,z)\} \; *_{2D} \; sPSF(x,y,z) \} \; *_{3D} \; emPSF(x,y,z)  \quad \quad(eq1)\] 
Finally, we get the image on the detector, by selecting the z-plane that corresponds to the focal plane ($z_{focal}$) from the last output cube as shown in the Figure \ref{fig:fwd_model1}. Then the noise is added to the resulting image using the following noise model discussed in the "EMCCD noise model" section.
\[\hat{Y}_t(x,y) = f_{EM} (\sim Poiss(Y_t(x,y,z_{focal}) + \sim Poiss(\bar{Y}_{Dark}) )+\sim Normal(0,\sigma_{Read})\quad \quad(eq2)\]
Here, $\sim Poiss(\mu)$ denotes the observations drawn from a Poisson distribution of mean $\mu$; $\sim Norm(\mu,\sigma)$ denotes the observations drawn from a Normal distribution of mean $\mu$ and standard deviation $\sigma$; $\bar{Y}_{Dark}$ is the expected value of dark current of the camera and $f_{EM}(.)$ models the electron multiplying process of the EMCCD camera. Figure \ref{fig:fwd_model2} shows the cross sections of the simulated PSFs and a representative simulated DEEP-TFM images.

For the beads experiments, fully simulated objects were used as $X0$s. For Neuronal experiments, experimentally acquired PSTPM images were used as $X0$s after pre-processing them to match the voxel size of the DEEP-TFM experiment. During training, the PSTPM images were simulated and used as the output (or the ground truth) of the network. The object $X0$ was convolved with the excitation PSF to generate the PSTPM image.

\begin{figure}[!ht]
\centering
\includegraphics[width=8 cm]{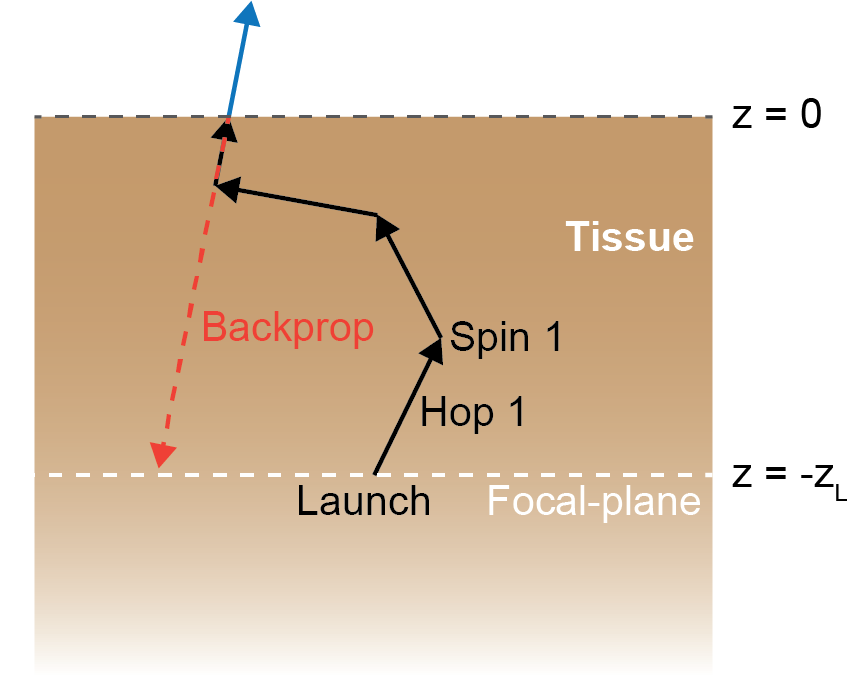}
\caption{Illustration of the light scattering process in a scattering tissue.}
\label{fig:sPSF}
\end{figure}

\paragraph{Scattering point spread function (sPSF).}In this section we discuss the mathematical model of the scattering point spread function. We define the scattering point spread function as the effective photon distribution of a certain z-plane below the tissue surface, due to a point source placed at that z-plane. Using Monte Carlo simulations of light transport through scattering tissue from a point source, we model how the down-stream optics see the light distribution from the plane where the point source is located. Figure \ref{fig:sPSF} Illustrates how the light scattering process is modeled using Mote Carlo methods. The surface is treated as the $z=0$ axial plane. A fluorescent light source is placed at $z=-z_0$ plane. The lateral location of this point source is considered as $x=y=0$. A photon is then launched from the source, placed at $(0,0,-z_0)$, in the random direction, $(u_x,u_y,u_z)$. This photon is propagated for s distance (call this a hop). Then a scattering event occurs, and the direction of the photon is changed by a deflection angle, $\theta$, and an azimuth angle, $\psi$ (call this a spin). Then the photon is hopped and spun again and again until it reaches the surface. Once it reaches the surface, the photon’s final direction is recorded. If the direction is within the range captured by the numerical aperture of the objective, the photon is then backpropagated until it reaches the original launch plane (i.e. $z=-z_0$) and its location on the launch plane is recorded. When the imaging system is focused at this launch plane the photon appears to have originated from this location. This process is related a sufficiently large number of times to generate the sPSF at the launch plane.

In the simulation we need to randomly generate the hop distance (s) and spin angles ($\theta$ and $\psi$) according to the scattering properties of the tissue. First, the probability distribution of s is related to the scattering coefficient of the tissue, $\mu_s$. By the definition of $\mu_s$ [ref], the probability of transmission of a photon without encountering a scattering event after path-length $s$ is given by,
\[P(S>s)=e^{-\mu_s s}\quad \quad(eq3)\]
Therefore, can derive the cumulative distribution function (c.d.f) of $s$, $F(s)$, and probability density function (p.d.f) of $s$, $f(s)$ as,
\[F(s)=P(S \leqslant s)=1-e^{-\mu_s s}\quad \quad(eq4)\]
\[f(s)=\frac{d(F(s))}{dt} = \mu_s e^{-\mu_s s}\quad \quad(eq5)\]
In Mote Carlo methods, $f(s)$ is sampled using a computer generated, uniformly distributed random number $rnd_1$ such that,
\[rnd_1=F(s) \quad \quad(eq6)\]
Therefore, we can derive $s$ as a function of $rnd_1$ as,
\[s=-ln(1-rnd_1)/\mu_s  \quad \quad(eq7)\]
Second, the deflection angle, $\theta$, is related to the anisotropy, $g$, of the tissue according to the Henyey-Greenstein scattering function (HG function) [ref] that mimics the angular dependence of light scattering by small particles [ref]. According to the HG function, the p.d.f of $cos(\theta)$ can be written as,
\[f(cos(\theta))=\frac{1}{2} \frac{(1-g^2)}{(1+g^2-2g \; cos(\theta))^{(3/2)}}   \quad \quad(eq8)\]
From the same principles of sampling $f(cos(\theta))$ using a second random number $rnd_2$ it can be shown that,
\[cos(\theta)=\frac{1}{2g} (1+g^2-(\frac{1-g^2}{1-g+2g \; rnd_2})^2 ) \quad \quad(eq9)\]
Last, the azimuth angle, $\psi$, can be picked at random, i.e. $\psi=rnd_3$, where $rnd_3$ is a third computer generated normally distributed random number. For more details we refer interested readers to Jacques and Wang \cite{jacques1995monte}.

\paragraph{EMCCD noise model.} Next we need to model the noise added to the image as shown in Eq2. We use an electron multiplication CCD (EMCCD) camera for detection. Let $Y_{Shot}$ be the number of electrons $\bar{e}$ from the signal detected before the electron multiplying (EM) process. Note that the expected value of $Y_{Shot}$ (i.e. $\bar{Y}_{Shot}$) is equal to  $Y_t(x,y,z_{focal})$ in Eq2. Let $Y_{Dark}$ be the number of electrons generated by the dark current of the camera. Dark current is usually listed in the camera specifications in "$\bar{e}/pixel/s$" and can be used to calculate the expected value of $Y_{Dark}$ (i.e. $\bar{Y}_{Dark}$) when the exposure time is known. Both $Y_{Shot}$ and $Y_{Dark}$ are Poisson distributed. Let, $N_{Read}$ be the read noise of the camera. The standard deviation of the read noise ($\sigma_{Read}$) is listed in camera specifications in $\bar{e}$. Read noise can be described by a normal distribution with zero mean and ($\sigma_{Read}$) standard deviation. Let’s denote the EM process by a statistical function $f_{EM}(.)$. Then we can write a formulation for the output signal, $Y_{Out}$ ($=\hat{Y}_t(x,y)$ in Eq2) , in $\bar{e}$ as,
\[Y_{Out} = f_{EM} (Y_{Shot}+Y_{Dark} )+N_{Read}\quad \quad(eq10)\]
We can then write the mean and variation equations for the signals from eq10 as,
\[ \bar{Y}_{Out} = g_{EM}(\bar{Y}_{Shot} + \bar{Y}_{Dark})\quad \quad(eq11) \]
\[\sigma_{Out}^2 = g_{EM}^2  F^2  (\sigma_{Shot}^2+\sigma_{Dark}^2 )+\sigma_{Read}^2\quad \quad(eq12)\]
Here $\bar{Y}$ denotes the mean of the random variable $Y$. $g_{EM}$ is the EM-gain of the camera and it is the average gain added by the EM process. The range of $g_{EM}$ is listed in camera specs and the exact value is set during imaging. $F$ quantifies the noise added by the EM process, $f_{EM}(.)$, and is discussed in the next subsection.



\paragraph{Electron Multiplication Process, $f_{EM}(.)$.}The EM process in practice is noisy and $g_{EM}$ is only the average value of the EM-gain. This noise is quantified by the excess noise factor (ENF), $F$, defined in Eq12. We can derive an expression for $F^2$ from Eq12 as,
\[ F^2=\frac{\sigma_{Out}^2-\sigma_{Read}^2}{g_{EM}^2  (\sigma_{Shot}^2+\sigma_{Dark}^2 )} \quad \quad(eq13) \]
Here $(\sigma_{Shot}^2+\sigma_{Dark}^2 )$ is the variation of the signal that goes into the EM process, and $(\sigma_{Out}^2-\sigma_{Read}^2)$ is the variation of the signal that comes out of the EM process (if F is 1 there is no noise added to by the EM process). Robins and Hadwen\cite{robbins2003noise} derived an expression for $F$ as,
\[ F^2= \frac{1}{g_{EM}} (\frac{2g_{EM}+\alpha-1}{\alpha+1}) = 2(g_{EM}-1) g_{EM}^{(-(N+1)/N)}+\frac{1}{g_{EM}} \quad \quad(eq14) \]
Here, $N$ is the number of EM gain stages. EM process is treated as Bernoulli process at each EM stage and $\alpha$ is the probability of an electron multiplication event happening, i.e. the probability of a success event. The value of α is usually small and is in the order of 1-2\% [ref- Nuvu cam data sheet]. Considering all input electrons to a stage, the gain for each EM stage can be described by a Binomial distribution with a probability mass function,
\[ pmf(x_{Ad};x_{in},\alpha) = Pr(X_{Ad}=x_{Ad}) = { {x_{in}}\choose{x_{Ad}} } \alpha^{x_{Ad}} (1-\alpha)^{(x_{in}-x_{Ad})} \quad \quad(eq15) \]
Here $x_{in}$ is the number of input electrons to the gain stage and $X_{Ad}$ is the number of added electrons through the EM gain stage. The output electrons from the gain therefore equal to, $X_{Ad}+x_{in}$. $N$ such gains stages are cascaded to get the final EM output signal $f_{EM}(.)$. 

However, simulating a cascade of $pmf(.)$ functions of the form Eq15 to add noise to every pixel in every simulated image is prohibitively slow. Therefore for each potential input value to $f_{EM}(.)$ we generated a distribution of $f_{EM}(.)$ output values. Note that this is doable as the inputs to $f_{EM}(.)$ are integers that come from Poisson distributed variables (i.e. $Y_{Shot}+Y_{Dark}$). Then during the simulations, we randomly sampled the output distribution corresponding to the input value to get the output of $f_{EM}(.)$ in Eq10.


\subsection*{Inverse Model}

\begin{figure}[!ht]
\centering
\includegraphics[width=\linewidth]{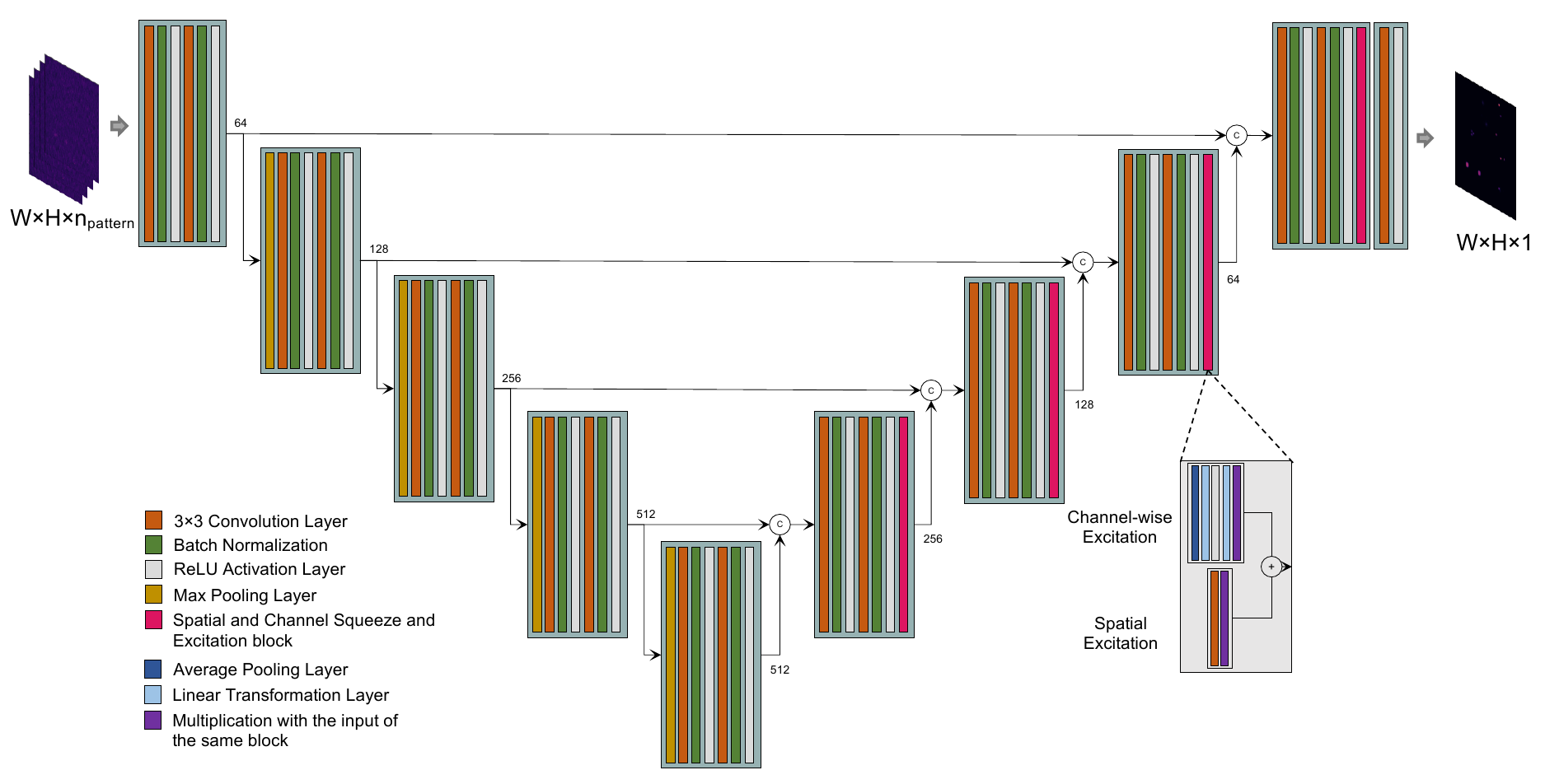}
\caption{Our proposed architecture: UNet with concurrent channel and spatial attention mechanism}
\label{fig:model}
\end{figure}


\paragraph{Network Architecture.} In this work, we use a deep learning network, called DEEP$^2$, as the inverse model. Our  proposed DEEP$^2$ is inspired by the  UNet\cite{ronneberger2015u} architecture with some modifications as illustrated in Figure \ref{fig:model}. DEEP$^2$ consists of 8 convolution blocks, each block containing 2 convolutional layers followed by batch normalization and a ReLU activation. In the encoder block, output of each block is sent through a max pooling layer for downsampling. In the decoder blocks, an additional block, which we call scSE, is integrated. The importance of this block is discussed in the next section. The skip connections between the downsample and upsample parts, enable the model to learn the fine grained details in the biological structures more efficiently. We exploit a separate block at the end, which consists of a convolutional layer and a ReLU activation layer, as the final reconstruction step. 

\paragraph{UNet with Spatial and channel squeeze and excitation.} In any deep learning model, the main purpose of convolution layer is to learn to capture all local spatial information within each channel and to create feature maps jointly encoding the spatial and channel dependencies. To improve joint encoding of spatial and channel information through convolution, it requires much effort. Thus, Concurrent Spatial and Channel ‘Squeeze and Excitation’ (scSE) block is introduced\cite{roy2018concurrent}. This improves the feature map by suppressing the weak features and  enhancing the important features. Despite, recalibrating non-significant features towards zero with scSE may increase the sparsity in deeper layers as well as reduce the parameter learning \cite{uhrig2017sparsity}. We denote the performance enhancement of the UNet with scSE over the vanilla UNet in our experiments. 

\paragraph{Loss Function.} Our learning algorithms focus on solving the optimization problem given by structural risk minimization, 
$\min _{f} \frac{1}{N} \sum_{i=1}^{N} L_{\theta}(I_{f}\hat{Y}_{t},X0)+\lambda R(I_{f})$, where $\hat{Y}_{t}$ is the set of DEEP-TFM images given to the Inverse model $I_{f}$ and the $X0$ is the corresponding ground truth (unscattered) image. The first term is the empirical risk, which is minimized while in the second term, $R(I_{f})$ i.e. regularization term,  represents the model complexity and $\lambda$ balances both terms. In the past few years, the attention of the Machine Learning community is drawn by the loss function, which provides a significant theoretical and practical value to the optimization task. Therefore, we have exploited 4 loss functions with the above described architecture as the inverse model to obtain the optimal learning algorithm. 
The least squared error given by Equation \ref{eq:mse}  is the most frequently used loss function in regression tasks. Let us assume the output of the inverse model is $\hat{X0}$,
\begin{equation}
Loss_{MSE} = \frac{1}{n} \sum_{t=1}^{n} \left(X0- \hat{X0}\right)^{2} \label{eq:mse}
\end{equation}
This loss function can converge fast as the gradient of the loss decreases when the loss approach 0. However, a drawback associated with this loss function is the high sensitivity for outliers. Root Mean Squared Logarithmic Error (RMSLE) is more robust to the outliers as it nullifies the effect by considering the $\log$ value. 
\begin{equation}
Loss_{RMSLE}=\sqrt{\frac{1}{n} \sum_{t=1}^{n}\left(\log \left(X0\right)-\log \left(\hat{X0}\right)\right)^{2}}, \label{eq:rmsle}
\end{equation}
Therefore, the above loss function is also utilized in our experiments. Next we also explored the Smooth L1 loss, which is also more robust for the outliers compared the mean squared error. 
\begin{equation}
Loss_{Smooth L1}=\left\{\begin{array}{ll}
\frac{1}{n} \sum_{t=1}^{n}\frac{1}{2}\left(X0-\hat{X0}\right)^{2}  & \text { if } \frac{1}{n} \sum_{t=1}^{n}\left|X0-\hat{X0}\right|< 1\\
\frac{1}{n} \sum_{t=1}^{n} \left|X0-\hat{X0}\right|-\frac{1}{2} , & \text { otherwise }
\end{array}\right.
\end{equation}
Smooth L1 loss conditionally chooses either the mean squared loss or the absolute loss, while combining the advantages of both loss functions. Nonetheless, we explored the usability of KL Divergence as a loss function for our inverse algorithm. 

\begin{equation}
Loss_{KL Divergence}=\frac{1}{n} \sum_{t=1}^{n} X0 \cdot\log \left(\frac{X0}{\hat{X0}}\right)
\end{equation}

KL divergence is identified as a loss function, which can handle distributions more efficiently despite the normality of the approximation error. 

\subsubsection*{Training and Evaluating the Inverse model} 

Learning rate is 0.001 for training the inverse model, while the momentum and weight decay are 0.9 and 0.0001 respectively. All the experiments are conducted with the PyTorch \cite{paszke2017automatic} learning framework. All the experiments are done with the Titan RTX available at FAS cluster. 

\paragraph{Evaluation Metrics.} We mainly used three metrics to evaluate the performance of our model on the testing data:(1)Structural Similarity Index (SSIM),(2)Mean Squared Error (MSE),(3)Peak Signal to Noise Ratio (PSNR)

\subsection*{Training and Validation Datasets} 

\paragraph{Artificial Fluorescent Beads Mixture Dataset.}For testing our framework, we have an experimentally acquired fluorescent beads instance using the DEEP-TFM. In addition, we require a large number of instances to optimize the learning based inverse model. Therefore, Artificial beads data, in the size of the experimentally acquired DEEP-TFM instance, were generated to use as the input to the forward model. The 3D volumes with beads in radius 1.5 um and 6 um were generated along with randomly selected intensities from a normal distribution with mean 1 and standard deviation 0.1. The 1.5 radius beads intensities were enhanced by 5 times compared to the 6 um beads. These artificially generated 3D volumes with beads were given as the input to the forward model to generate the synthetic DEEP-TFM instances. Since the experimental beads data acquired using DEEP-TFM was obtained at 5 scattering lengths, the same depth had been used for the synthetic bead dataset creation. Thus, for 5 scattering length, 3455 instances of were obtained which we split into 4:1 ratio for training and validation. In addition, 128 artificial beads instances were generated for independent testing. Each instance consists of 32 patterned DEEP-TFM like images of size 256 pixels by 256 pixels. 

\paragraph{Mouse Pyramidal Neuron Dataset.} Experimentally obtained images of pyramidal neurons with the dendritic arbor, using a PSTPM, were also used for synthetic DEEP-TFM image generation using the forward model. Each neuronal image volume was 607 pixels * 607 pixels * 263 pixels with a spatial resolution of 0.25 um. We used 11 such neuronal image volumes for creation of synthetic DEEP-TFM neuronal instances. In order to enhance the intensity of the fine structures such as dendritic spines, the intensity was thresholded at 20, by assigning 20 for all the pixels with intensity over 20. The successive images along the z plane of the image volume were containing similar structures. Thus, to avoid repeating similar training instances,  the sub image volumes that matches the third dimension of the exPSF, were extracted from the neuronal volume, to use as the input to the forward model, with 5 image plane apart from each other. Further, to avoid the image volumes with no significant structures, the sub image volumes with the mean exceeding the mean of the whole neuronal image volume were considered as the input of the forward model. The 3895 synthetic DEEP-TFM neuronal instances were generated for 2 (~100 um) and 6 (~300 um) scattering lengths. Each instance consists of 326 pixels by 326 pixels ground truth ($\hat{X}_{0}(x, y)$) and 32 patterned images ($t$ = 32) of size 326 pixels by 326 pixels ($\hat{Y}_{t}(x, y)$) as the input of the inverse model $I$. 

\paragraph{Mouse Cortical Vasculature Dataset.}In addition to the experimentally acquired beads instances, we also acquired the blood vessel instances at increasing depths using the DEEP-TFM. Hence, to train the network, we used an equivalent vascular structural image volume acquired using a three-photon Fluorescent microscope \cite{yildirim2020quantitative}. The vascular image stack consists of 800 z-stacks, each had an increment of 1.5 um and each sized 512 pixels * 512 pixels with a spatial resolution of 0.6um corresponding to a field of view of ~300um. 

\paragraph{Artificial Vasculature Dataset.}
Additionally, artificial blood vessel volumes were employed \cite{schneider2012tissue,todorov2020machine} to demonstrate the applicability of this framework on vasculature structures furthermore. The synthetic volumes originally consisted of blood vessels with radius ranging from 1-7 pixels. The finest blood vessels with low intensities were removed by applying a threshold at 190. The blood vessel volumes were originally sized 325 pixels by 304 pixels by 600 pixels and 20 such volumes were downloaded. Each volume was rescaled by factor 1.09, normalized by dividing by the maximum intensity of each volume and extracted volumes of 326 pixels by 326 pixels by 200 pixels volumes to redirect as the input to the Forward model. Similarly, we generated datasets at 2 and 6 scattering lengths for blood vessels. 


\section*{Acknowledgements}
This work was supported by the Center for Advanced Imaging at Harvard University (DNW, NW, MA), 5-P41EB015871-32 (DNW, PS), R21 MH130067 (PS, DNW), R21 NS105070 (PS), and R00EB027706 (MY). DNW is also supported by the John Harvard Distinguished Science Fellowship Program within the FAS Division of Science of Harvard University. We also thank Dr. Josiah R. Boivin for providing point scanning two-photon datasets of pyramidal neurons.

\bibliography{sample}

\begin{thebibliography}{10}
\urlstyle{rm}
\expandafter\ifx\csname url\endcsname\relax
  \def\url#1{\texttt{#1}}\fi
\expandafter\ifx\csname urlprefix\endcsname\relax\def\urlprefix{URL }\fi
\expandafter\ifx\csname doiprefix\endcsname\relax\def\doiprefix{DOI: }\fi
\providecommand{\bibinfo}[2]{#2}
\providecommand{\eprint}[2][]{\url{#2}}

\bibitem{rocheleau2003two}
\bibinfo{author}{Rocheleau, J.~V.} \& \bibinfo{author}{Piston, D.~W.}
\newblock \bibinfo{journal}{\bibinfo{title}{Two-photon excitation microscopy
  for the study of living cells and tissues}}.
\newblock {\emph{\JournalTitle{Current protocols in cell biology}}}
  \textbf{\bibinfo{volume}{20}}, \bibinfo{pages}{4--11} (\bibinfo{year}{2003}).

\bibitem{yildirim2019functional}
\bibinfo{author}{Yildirim, M.}, \bibinfo{author}{Sugihara, H.},
  \bibinfo{author}{So, P.~T.} \& \bibinfo{author}{Sur, M.}
\newblock \bibinfo{journal}{\bibinfo{title}{Functional imaging of visual
  cortical layers and subplate in awake mice with optimized three-photon
  microscopy}}.
\newblock {\emph{\JournalTitle{Nature communications}}}
  \textbf{\bibinfo{volume}{10}}, \bibinfo{pages}{1--12} (\bibinfo{year}{2019}).

\bibitem{yildirim2022label}
\bibinfo{author}{Yildirim, M.} \emph{et~al.}
\newblock \bibinfo{journal}{\bibinfo{title}{Label-free three-photon imaging of
  intact human cerebral organoids for tracking early events in brain
  development and deficits in rett syndrome}}.
\newblock {\emph{\JournalTitle{eLife}}} \textbf{\bibinfo{volume}{11}},
  \bibinfo{pages}{e78079} (\bibinfo{year}{2022}).

\bibitem{benninger2013two}
\bibinfo{author}{Benninger, R.~K.} \& \bibinfo{author}{Piston, D.~W.}
\newblock \bibinfo{journal}{\bibinfo{title}{Two-photon excitation microscopy
  for the study of living cells and tissues}}.
\newblock {\emph{\JournalTitle{Current protocols in cell biology}}}
  \textbf{\bibinfo{volume}{59}}, \bibinfo{pages}{4--11} (\bibinfo{year}{2013}).

\bibitem{vaziri2010ultrafast}
\bibinfo{author}{Vaziri, A.} \& \bibinfo{author}{Shank, C.~V.}
\newblock \bibinfo{journal}{\bibinfo{title}{Ultrafast widefield optical
  sectioning microscopy by multifocal temporal focusing}}.
\newblock {\emph{\JournalTitle{Optics express}}} \textbf{\bibinfo{volume}{18}},
  \bibinfo{pages}{19645--19655} (\bibinfo{year}{2010}).

\bibitem{rowlands2017wide}
\bibinfo{author}{Rowlands, C.~J.} \emph{et~al.}
\newblock \bibinfo{journal}{\bibinfo{title}{Wide-field three-photon excitation
  in biological samples}}.
\newblock {\emph{\JournalTitle{Light: Science \& Applications}}}
  \textbf{\bibinfo{volume}{6}}, \bibinfo{pages}{e16255--e16255}
  (\bibinfo{year}{2017}).

\bibitem{escobet2018wide}
\bibinfo{author}{Escobet-Montalb{\'a}n, A.} \emph{et~al.}
\newblock \bibinfo{journal}{\bibinfo{title}{Wide-field multiphoton imaging
  through scattering media without correction}}.
\newblock {\emph{\JournalTitle{Science advances}}}
  \textbf{\bibinfo{volume}{4}}, \bibinfo{pages}{eaau1338}
  (\bibinfo{year}{2018}).

\bibitem{wadduwage2019scattering}
\bibinfo{author}{Zheng, C.} \emph{et~al.}
\newblock \bibinfo{journal}{\bibinfo{title}{De-scattering with excitation
  patterning enables rapid wide-field imaging through scattering media}}.
\newblock {\emph{\JournalTitle{Science Advances}}}
  \textbf{\bibinfo{volume}{7}}, \bibinfo{pages}{eaay5496}
  (\bibinfo{year}{2021}).

\bibitem{belthangady2019applications}
\bibinfo{author}{Belthangady, C.} \& \bibinfo{author}{Royer, L.~A.}
\newblock \bibinfo{journal}{\bibinfo{title}{Applications, promises, and
  pitfalls of deep learning for fluorescence image reconstruction}}.
\newblock {\emph{\JournalTitle{Nature methods}}} \bibinfo{pages}{1--11}
  (\bibinfo{year}{2019}).

\bibitem{weigert2018content}
\bibinfo{author}{Weigert, M.} \emph{et~al.}
\newblock \bibinfo{journal}{\bibinfo{title}{Content-aware image restoration:
  pushing the limits of fluorescence microscopy}}.
\newblock {\emph{\JournalTitle{Nature methods}}} \textbf{\bibinfo{volume}{15}},
  \bibinfo{pages}{1090--1097} (\bibinfo{year}{2018}).

\bibitem{jin2017deep}
\bibinfo{author}{Jin, K.~H.}, \bibinfo{author}{McCann, M.~T.},
  \bibinfo{author}{Froustey, E.} \& \bibinfo{author}{Unser, M.}
\newblock \bibinfo{journal}{\bibinfo{title}{Deep convolutional neural network
  for inverse problems in imaging}}.
\newblock {\emph{\JournalTitle{IEEE Transactions on Image Processing}}}
  \textbf{\bibinfo{volume}{26}}, \bibinfo{pages}{4509--4522}
  (\bibinfo{year}{2017}).

\bibitem{krizhevsky2012imagenet}
\bibinfo{author}{Krizhevsky, A.}, \bibinfo{author}{Sutskever, I.} \&
  \bibinfo{author}{Hinton, G.~E.}
\newblock \bibinfo{title}{Imagenet classification with deep convolutional
  neural networks}.
\newblock In \emph{\bibinfo{booktitle}{Advances in neural information
  processing systems}}, \bibinfo{pages}{1097--1105} (\bibinfo{year}{2012}).

\bibitem{ziletti2018insightful}
\bibinfo{author}{Ziletti, A.}, \bibinfo{author}{Kumar, D.},
  \bibinfo{author}{Scheffler, M.} \& \bibinfo{author}{Ghiringhelli, L.~M.}
\newblock \bibinfo{journal}{\bibinfo{title}{Insightful classification of
  crystal structures using deep learning}}.
\newblock {\emph{\JournalTitle{Nature communications}}}
  \textbf{\bibinfo{volume}{9}}, \bibinfo{pages}{1--10} (\bibinfo{year}{2018}).

\bibitem{wei2019dominant}
\bibinfo{author}{Wei, Z.}, \bibinfo{author}{Liu, D.} \& \bibinfo{author}{Chen,
  X.}
\newblock \bibinfo{journal}{\bibinfo{title}{Dominant-current deep learning
  scheme for electrical impedance tomography}}.
\newblock {\emph{\JournalTitle{IEEE Transactions on Biomedical Engineering}}}
  \textbf{\bibinfo{volume}{66}}, \bibinfo{pages}{2546--2555}
  (\bibinfo{year}{2019}).

\bibitem{liu2014early}
\bibinfo{author}{Liu, S.} \emph{et~al.}
\newblock \bibinfo{title}{Early diagnosis of alzheimer's disease with deep
  learning}.
\newblock In \emph{\bibinfo{booktitle}{2014 IEEE 11th international symposium
  on biomedical imaging (ISBI)}}, \bibinfo{pages}{1015--1018}
  (\bibinfo{organization}{IEEE}, \bibinfo{year}{2014}).

\bibitem{wang2016accelerating}
\bibinfo{author}{Wang, S.} \emph{et~al.}
\newblock \bibinfo{title}{Accelerating magnetic resonance imaging via deep
  learning}.
\newblock In \emph{\bibinfo{booktitle}{2016 IEEE 13th International Symposium
  on Biomedical Imaging (ISBI)}}, \bibinfo{pages}{514--517}
  (\bibinfo{organization}{IEEE}, \bibinfo{year}{2016}).

\bibitem{girshick2014rich}
\bibinfo{author}{Girshick, R.}, \bibinfo{author}{Donahue, J.},
  \bibinfo{author}{Darrell, T.} \& \bibinfo{author}{Malik, J.}
\newblock \bibinfo{title}{Rich feature hierarchies for accurate object
  detection and semantic segmentation (pp. 580--587)}.
\newblock In \emph{\bibinfo{booktitle}{CVPR’14: Proceedings of the 2014 IEEE
  Conference on Computer Vision and Pattern Recognition, IEEE Computer
  Society}} (\bibinfo{year}{2014}).

\bibitem{nielsen2018deep}
\bibinfo{author}{Nielsen, A.~A.} \& \bibinfo{author}{Voigt, C.~A.}
\newblock \bibinfo{journal}{\bibinfo{title}{Deep learning to predict the
  lab-of-origin of engineered dna}}.
\newblock {\emph{\JournalTitle{Nature communications}}}
  \textbf{\bibinfo{volume}{9}}, \bibinfo{pages}{1--10} (\bibinfo{year}{2018}).

\bibitem{eraslan2019single}
\bibinfo{author}{Eraslan, G.}, \bibinfo{author}{Simon, L.~M.},
  \bibinfo{author}{Mircea, M.}, \bibinfo{author}{Mueller, N.~S.} \&
  \bibinfo{author}{Theis, F.~J.}
\newblock \bibinfo{journal}{\bibinfo{title}{Single-cell rna-seq denoising using
  a deep count autoencoder}}.
\newblock {\emph{\JournalTitle{Nature communications}}}
  \textbf{\bibinfo{volume}{10}}, \bibinfo{pages}{1--14} (\bibinfo{year}{2019}).

\bibitem{eulenberg2017reconstructing}
\bibinfo{author}{Eulenberg, P.} \emph{et~al.}
\newblock \bibinfo{journal}{\bibinfo{title}{Reconstructing cell cycle and
  disease progression using deep learning}}.
\newblock {\emph{\JournalTitle{Nature communications}}}
  \textbf{\bibinfo{volume}{8}}, \bibinfo{pages}{1--6} (\bibinfo{year}{2017}).

\bibitem{wei2019physics}
\bibinfo{author}{Wei, Z.} \& \bibinfo{author}{Chen, X.}
\newblock \bibinfo{journal}{\bibinfo{title}{Physics-inspired convolutional
  neural network for solving full-wave inverse scattering problems}}.
\newblock {\emph{\JournalTitle{IEEE Transactions on Antennas and Propagation}}}
  \textbf{\bibinfo{volume}{67}}, \bibinfo{pages}{6138--6148}
  (\bibinfo{year}{2019}).

\bibitem{zhu2018image}
\bibinfo{author}{Zhu, B.}, \bibinfo{author}{Liu, J.~Z.},
  \bibinfo{author}{Cauley, S.~F.}, \bibinfo{author}{Rosen, B.~R.} \&
  \bibinfo{author}{Rosen, M.~S.}
\newblock \bibinfo{journal}{\bibinfo{title}{Image reconstruction by
  domain-transform manifold learning}}.
\newblock {\emph{\JournalTitle{Nature}}} \textbf{\bibinfo{volume}{555}},
  \bibinfo{pages}{487--492} (\bibinfo{year}{2018}).

\bibitem{wei20193d}
\bibinfo{author}{Wei, Z.} \emph{et~al.}
\newblock \bibinfo{journal}{\bibinfo{title}{3d deep learning enables fast
  imaging of spines through scattering media by temporal focusing microscopy}}.
\newblock {\emph{\JournalTitle{arXiv preprint arXiv:2001.00520}}}
  (\bibinfo{year}{2019}).

\bibitem{jacques1995monte}
\bibinfo{author}{Jacques, S.~L.} \& \bibinfo{author}{Wang, L.}
\newblock \bibinfo{title}{Monte carlo modeling of light transport in tissues}.
\newblock In \emph{\bibinfo{booktitle}{Optical-thermal response of
  laser-irradiated tissue}}, \bibinfo{pages}{73--100}
  (\bibinfo{publisher}{Springer}, \bibinfo{year}{1995}).

\bibitem{robbins2003noise}
\bibinfo{author}{Robbins, M.~S.} \& \bibinfo{author}{Hadwen, B.~J.}
\newblock \bibinfo{journal}{\bibinfo{title}{The noise performance of electron
  multiplying charge-coupled devices}}.
\newblock {\emph{\JournalTitle{IEEE transactions on electron devices}}}
  \textbf{\bibinfo{volume}{50}}, \bibinfo{pages}{1227--1232}
  (\bibinfo{year}{2003}).

\bibitem{ronneberger2015u}
\bibinfo{author}{Ronneberger, O.}, \bibinfo{author}{Fischer, P.} \&
  \bibinfo{author}{Brox, T.}
\newblock \bibinfo{title}{U-net: Convolutional networks for biomedical image
  segmentation}.
\newblock In \emph{\bibinfo{booktitle}{International Conference on Medical
  image computing and computer-assisted intervention}},
  \bibinfo{pages}{234--241} (\bibinfo{organization}{Springer},
  \bibinfo{year}{2015}).

\bibitem{roy2018concurrent}
\bibinfo{author}{Roy, A.~G.}, \bibinfo{author}{Navab, N.} \&
  \bibinfo{author}{Wachinger, C.}
\newblock \bibinfo{title}{Concurrent spatial and channel ‘squeeze \&
  excitation’in fully convolutional networks}.
\newblock In \emph{\bibinfo{booktitle}{International Conference on Medical
  Image Computing and Computer-Assisted Intervention}},
  \bibinfo{pages}{421--429} (\bibinfo{organization}{Springer},
  \bibinfo{year}{2018}).

\bibitem{uhrig2017sparsity}
\bibinfo{author}{Uhrig, J.} \emph{et~al.}
\newblock \bibinfo{title}{Sparsity invariant cnns}.
\newblock In \emph{\bibinfo{booktitle}{2017 international conference on 3D
  Vision (3DV)}}, \bibinfo{pages}{11--20} (\bibinfo{organization}{IEEE},
  \bibinfo{year}{2017}).

\bibitem{paszke2017automatic}
\bibinfo{author}{Paszke, A.} \emph{et~al.}
\newblock \bibinfo{journal}{\bibinfo{title}{Automatic differentiation in
  pytorch}}.
\newblock {\emph{\JournalTitle{-}}}  (\bibinfo{year}{2017}).

\bibitem{yildirim2020quantitative}
\bibinfo{author}{Yildirim, M.} \emph{et~al.}
\newblock \bibinfo{journal}{\bibinfo{title}{Quantitative third-harmonic
  generation imaging of mouse visual cortex areas reveals correlations between
  functional maps and structural substrates}}.
\newblock {\emph{\JournalTitle{Biomedical Optics Express}}}
  \textbf{\bibinfo{volume}{11}}, \bibinfo{pages}{5650--5673}
  (\bibinfo{year}{2020}).

\bibitem{schneider2012tissue}
\bibinfo{author}{Schneider, M.}, \bibinfo{author}{Reichold, J.},
  \bibinfo{author}{Weber, B.}, \bibinfo{author}{Sz{\'e}kely, G.} \&
  \bibinfo{author}{Hirsch, S.}
\newblock \bibinfo{journal}{\bibinfo{title}{Tissue metabolism driven arterial
  tree generation}}.
\newblock {\emph{\JournalTitle{Medical image analysis}}}
  \textbf{\bibinfo{volume}{16}}, \bibinfo{pages}{1397--1414}
  (\bibinfo{year}{2012}).

\bibitem{todorov2020machine}
\bibinfo{author}{Todorov, M.~I.} \emph{et~al.}
\newblock \bibinfo{journal}{\bibinfo{title}{Machine learning analysis of whole
  mouse brain vasculature}}.
\newblock {\emph{\JournalTitle{Nature methods}}} \textbf{\bibinfo{volume}{17}},
  \bibinfo{pages}{442--449} (\bibinfo{year}{2020}).

\end{thebibliography}

\end{document}